\documentclass[a4paper,11pt]{article}
\usepackage{jheppub} 
\usepackage[T1]{fontenc} 

\newcommand{\beqn}{\begin{eqnarray}}
\newcommand{\beq}{\begin{equation}}
\newcommand{\eeqn}{\end{eqnarray}}
\newcommand{\eeq}{\end{equation}}

\title{\boldmath Color confinement due to spontaneous breaking of magnetic $U(1)_m^8$}
\author{Tsuneo Suzuki}
\affiliation{Research Center for Nuclear Physics, Osaka University,\\Osaka 567-0047, Japan}
\emailAdd{tsuneo@rcnp.osaka-u.ac.j}

\abstract{It is possible to understand non-Abelian color confinement in the framework of the Abelian dual Meissner effect, if the non-Abelian Bianchi identity is violated. The violation is equal to  8 Abelian monopole currents of the Dirac type satisfying Abelian conservation  rules kinematically. There exist magnetic $U(1)_m^8$ symmetries in non-Abelian $SU(3)$ QCD. When the magnetic $U(1)_m^8$ symmetries are broken spontaneously, only states which are invariant under all $U(1)_e$ subgroups of $SU(3)$ can exist as a physical state. Such states are  just $SU(3)$ singlets. The QCD vacuum in the confinement phase is characterized by one long percolating monopole loop   
running over the whole lattice volume in both quenched and full QCD. The long loop in full QCD  is on average a few times longer in comparison with that in quenched QCD case at a similar lattice distance on the same volume.  Surprisingly, the monopole behaviors in  full QCD seem to be independent of the bare quark mass suggesting irrelevance of Abelian monopoles to the chiral symmetry breaking mechanism in QCD.  
  Existence of such Abelian magnetic monopoles in the continuum limit is studied in detail in $SU(3)$ by means of a block spin transformation of monopoles and the inverse Monte-Carlo method.  The monopole density $\rho$ and the infrared effective monopole action $S(k)$ of  $n$ blocked monopoles  are determined for  $a(\beta)=(0.04\sim 2)$fm and $n=1\sim 12$ blockings on $48^4$ lattice in quenched QCD and for $a(\beta)=0.0846(7)$fm and $n=1\sim 24$ on $96^4$ in full QCD at $m_\pi=146$MeV.   Originally $\rho$ and $S(k)$ are a two-point function of $a(\beta)$ and the number of times of the blocking transformation $n$. However, both  are  found to be a function of $b=na(\beta)$ alone in the quenched QCD which suggests the existence of the continuum limit. In the full QCD, the renormalization flow is observed similarly but the scaling is not yet proved since the vacuum ensembles  are only at $\beta=1.82$.  The distributions of the long loops show that monopole condensation occurs due to the entropy dominance over the energy for all $b$ considered. 
}


\begin{document} 
\maketitle
\flushbottom

\section{Introduction}
\label{sec:intro}
Color confinement in the presence of  light quarks and massless gluons in QCD is not yet clarified. The problem is known to be related to the Yang-Mills (YM) existence and mass gap problem\cite{CMI:2000mp}. It is very interesting to find  a fundamental mechanism behind these problems. The dual Meissner picture proposed by 'tHooft\cite{tHooft:1975pu} and Mandelstam\cite{Mandelstam:1974pi} is promising when color magnetic monopoles which condense in the vacuum are found naturally in QCD. However, it is theoretically  difficult to find such magnetic monopoles in QCD composed of color electric fields alone without any artificial assumption as done say in Ref.\cite{tHooft:1981ht}.

   In 2014, motivated by a work\cite{Bonati:2010tz} which found violation of non-Abelian Bianchi identity (VNABI) exists behind the 'tHooft Abelian monopoles\cite{tHooft:1981ht}, the present author\cite{Suzuki:2014wya,Suzuki:2017lco,Suzuki:20220422} found an interesting fact that violation of non-Abelian Bianchi identity (VNABI) if it occurs is equivalent to 8 Abelian magnetic monopole currents satisfying Abelian conservation $\partial_\mu k_\mu^a(x)=0 \ \ (a=1\sim 8)$ kinematically\cite{Arafune:1974uy}. VNABI can exist when original gauge fields contain a singularity of the Dirac type\cite{Dirac:1931}.  If condensation of such 8 magnetic monopoles occurs, any state which has an electric color charge with respect to any electric $U(1)_e$ subgroup in $SU(3)$ is confined due to the Abelian dual Meissner effect. States neutral under all $U(1)_e$ subgroups are $SU(3)$ singlets alone. Hence non-Abelian $SU(3)$ color confinement is proved through the Abelian dual Meissner mechanism due to the Abelian monopole condensation with respect to  $U(1)_m^8$.   

    We show again how VNABI leads us to existence of 8 Abelian monopole currents satisfying conservation rules\cite{Suzuki:2014wya}.    
Define a covariant derivative operator $D_{\mu}=\partial_{\mu}-igA_{\mu}$. The Jacobi identities are expressed as $\epsilon_{\mu\nu\rho\sigma}[D_{\nu},[D_{\rho},D_{\sigma}]]=0$.
By direct calculations, we get
$[D_{\rho},D_{\sigma}]=-igG_{\rho\sigma}+[\partial_{\rho},\partial_{\sigma}]$,
where the second commutator term of the partial derivative operators cannot be discarded in general, since gauge fields may contain a line singularity. Actually, it is the origin of the violation of the non-Abelian Bianchi identities (VNABI) as shown in the following. The non-Abelian Bianchi identities and the Abelian-like Bianchi identities are, respectively: $D_{\nu}G^{*}_{\mu\nu}=0$ and $\partial_{\nu}f^{*}_{\mu\nu}=0$, where $f_{\mu\nu}$ is defined as $f_{\mu\nu}=\partial_{\mu}A_{\nu}-\partial_{\nu}A_{\mu}=(\partial_{\mu}A^a_{\nu}-\partial_{\nu}A^a_{\mu})\lambda^a/2$.
The relation $[D_{\nu},G_{\rho\sigma}]=D_{\nu}G_{\rho\sigma}$ and the Jacobi identities lead us to
\begin{eqnarray}
D_{\nu}G^{*}_{\mu\nu}=-\frac{i}{2g}\epsilon_{\mu\nu\rho\sigma}[D_{\nu},[\partial_{\rho},\partial_{\sigma}]]
=\frac{1}{2}\epsilon_{\mu\nu\rho\sigma}[\partial_{\rho},\partial_{\sigma}]A_{\nu}
=\partial_{\nu}f^{*}_{\mu\nu}. \label{eq-JK}
\end{eqnarray}
 Namely Eq.(\ref{eq-JK}) shows that VNABI, if exists,  is equivalent to a set of Abelian-like Bianchi identities.
Denote VNABI as  $J_{\mu}=D_{\nu}G^*_{\mu \nu}$ and Abelian-like monopole currents $k_{\mu}\equiv k_\mu^a\lambda^a/2$ without any gauge-fixing as the violation of the Abelian-like Bianchi identities:
$k_{\mu}=\partial_{\nu}f^*_{\mu\nu}
=\frac{1}{2}\epsilon_{\mu\nu\rho\sigma}\partial_{\nu}f_{\rho\sigma}.$ 
Eq.(\ref{eq-JK}) shows that 
$J_{\mu}=k_{\mu}$.
The Abelian-like monopole currents satisfy an Abelian conservation rule kinematically, $\partial_\mu k_\mu(x)=0$\cite{Arafune:1974uy}. There  exist  
exact Abelian (but kinematical) $U(1)_m^8$ symmetries in non-Abelian QCD. However the existence of the above Abelian monopoles is not related to any topological property like a case in Ref.\cite{tHooft:1981ht} and hence the Dirac quantization condition\cite{Dirac:1931} is not shown a priori. It must be proved rigorously if such Abelian monopoles play an important role in nature.

\section{Non-Abelian color confinement due to spontaneous breaking of $U(1)_m^8$}
\label{sec:theory}
Now we show explicitly why the Abelian dual Meissner effects due to spontaneous breaking of the Abelian magnetic $U(1)_m^8$ from $\partial_\mu k_\mu^a=0 \ \ (a=1\sim 8)$ lead us to non-Abelian color confinement.  To say about the Abelian dual Meissner effect, it is necessary to define a local $U(1)_e$ symmetry having a photon-like field from $SU(3)$. An arbitrary $SU(3)$ gauge transformation is expressed as 
\begin{eqnarray}
V(x)=e^{i\sum_{i=1,8}\alpha^i(x)X^i},
\end{eqnarray}
where $X^i$ is a generator of $su(3)$ Lie algebra.
There are eight $U(1)_e^i$ subgroups expressed in terms of $\alpha^i$  alone among $SU(3)$, although they are not commutative unless they belong to the same maximal torus group $U(1)^2$. 
Consider for  example, one $U(1)_m^3$ corresponding to the monopole current conservation $\partial_\mu k_\mu^3(x)=0$, where $k_\mu^3(x)$ comes from 
the singularity of $A_\mu^3(x)$. Now consider a $U(1)_e^3$ subgroup where $A_\mu^3(x)$ plays a role of a photon. If the monopoles $k_\mu^3(x)$ condense in the vacuum, $U(1)_m^3$ is broken spontaneously. The Abelian dual Meissner effect due to the monopole condensation leads us to that only the neutral states $|H>$ under $U(1)_e^3$ can exist in nature, that is , 
\begin{eqnarray}
e^{i\alpha^3(x)X^3}|H> = |H>.
\end{eqnarray}
The same situations occur with respect to all color components similarly.
Hence physical states which are neutral with respect to any $U(1)_e^i$ satisfy $e^{i\alpha^i(x)X^i}|H> = |H>$ for $i=1\sim 8$. 
Then it is possible to show that
\begin{eqnarray}
\tilde{V}(x)|H> = |H>, \label{U1^8}
\end{eqnarray}
where $\tilde{V}(x)\equiv \Pi_{i=1}^8 e^{i\alpha^i(x)X^i}$ and the ordering of the product is fixed arbitrary. Note that $\tilde{V}(x)$ has eight parameters
 satisfying $\tilde{V}(x)\tilde{V}^{\dag}(x)=1$ and $det(\tilde{V}(x))=1$.
Hence $\tilde{V}(x)\in SU(3)$.  Eq.(\ref{U1^8}) indicates that 
the state $|H>$ is $SU(3)$ invariant. Hence the Abelian dual Meissner effect due to spontaneous breaking of magnetic $U(1)_m^8$ can explain non-Abelian color confinement of QCD. 
\begin{table}[tb]
\begin{center}
\begin{tabular}{c|c|c|c|c|c|c|c|c|c|c}
\hline\hline
$\beta$&$m_\pi$&color&n&maxlp&lpnum&lp4&n&maxlp&lpnum&lp4\\
\hline\hline
2.7&quench& 1&1&969444(420)&22620(22)&20440(21)&8&3652(10)&6(0)&5(0)\\
   &        & 3&1&1016422(212)&15446(13)&14623(13)&8&3809(11)&6(0)&5(0)\\
   \hline
1.9 &702(1)&1&1& 3914480(209)&8112(9)&8111(9)&8&9210(22)&6(0)&5(0)\\
    &      &3&1& 3914649(189)&8084(10)&8083(10)&8&9220(22)&6(0)&5(0)\\
    &384(3)&1&1& 3914572(183)&8100(11)&8098(11)&8&9237(22)&5(0)&4(0)\\
    &      &3&1& 3914908(173)&8083(10)&8082(10)&8&9262(18)&6(0)&5(0)\\    
    &156(6)&1&1& 3914682(188)&8079(9)& 8078(9)&8&9233(22)&6(0)&5(0)\\
    &      &3&1& 3915081(156)&8086(10)& 8085(10)&8&9217(18)&6(0)&5(0)\\
 \hline\hline
\end{tabular}
\caption{\label{looplength}
Loop length distribution for 80 original and blocked monopole configurations.  (The coupling constant $\beta$,  the pion mass, color, blocking number of times $n$, longest loop length (maxlp), total number of loops (lpnum) and number of loops with length=4 (lp4)). }
\end{center}
\end{table}
\begin{table}[tb]
\begin{center}
\begin{tabular}{c|c|c|c|c|c|c|c|c}
\hline\hline
$m_\pi$ &$n$&density&$n$&density&$n$&density&$n$&density \\
\hline
quench&1& 0.76337(47)&2&1.91557(123)&4 &4.11464(175)&8&8.2855(73)\\
\hline
702(1)& 1& 1.91166(2)&2&3.83134(23)&4&7.66345(129)&8&15.3113(127)\\
384(3)& 1 & 1.91173(2)&2 &3.83120(19)&4 &7.65998(141)&8 &15.3275(122)\\
156(6)& 1 & 1.91171(2)&2 &3.83130(25)&4 &7.66164(177)&8 &15.3300(120)\\
\hline\hline
\end{tabular}
\caption{\label{density}
Monopole density  of $\beta=2.7$ quenched and $\beta=1.9$ full  QCD on $32^3\times 64$. $m_\pi$ is the pion mass and $n$ is the  blocking number.}
\end{center}
\end{table}

\section{Monopole behaviors in the lattice QCD framework}
\label{sec.behavior}
It is very interesting to study whether the above scenario is realized in nature or not in the lattice QCD framework rigorously.

First we have to define an Abelian monopole on lattice. 
In compact QED, DeGrand and Toussaint(DGT)\cite{DeGrand:1980eq} defined Abelian monopoles on lattice on the basis of the Dirac quantization condition corresponding to Abelian monopoles of the Dirac type. A plaquette variable 
$\theta_{\mu\nu}(s)\equiv \partial_\mu\theta_\nu(s)-\partial_\nu\theta_\mu(s)$ where $\theta_\mu(s)$ is an Abelian link field is decomposed into 
$\theta_{\mu\nu}(s)= \bar{\theta}_{\mu\nu}(s) +2\pi n_{\mu\nu}(s)$,
 where $\bar{\theta}_{\mu\nu}(s)\in [-\pi,\pi]$. 
 Then the integer $n_{\mu\nu}(s)\in [-2,2]$ 
could be regarded as a number of the Dirac string penetrating the plaquette, although the number can take $[-\infty,\infty]$ in the continuum theory. Then DGT~\cite{DeGrand:1980eq} defined a monopole current on lattice as 
\begin{eqnarray}
k_{\mu}(s) = -\frac{1}{4\pi}\epsilon_{\mu\nu\rho\sigma}
\partial_{\nu}\bar{\theta}_{\rho\sigma}(s+\hat{\mu}). 
\label{dgcur} 
\end{eqnarray}
Since the Dirac quantization condition\cite{Dirac:1931} is essential for existence of Abelian monopoles in QCD, we adopt here in $SU(3)$ the above lattice Abelian monopole definition\cite{DeGrand:1980eq} for each global color  and  study the continuum limit. 

Now let us study monopole distributions both in full QCD and quenched $SU(3)$ QCD of the Iwasaki gauge action\cite{Iwasaki:1985,IKKY:1997,Takeda:2004}. The full $2+1$ flavor QCD configurations in this section are cited from the works done by PACS-CS collaboration\cite{Aoki:2009} using the nonperturbatively $O(a)$-improved Wilson quark action and Iwasaki gauge action on $32^3\times 64$ lattice. There exist six data with different pion masses ($m_\pi=702(1), 569(2), 411(2), 384(3), 295(3), 156(6)$MeV) at $\beta=1.9$   corresponding to the lattice spacing $a=0.0907(13)$ fm.  For comparison, we adopt quenched $SU(3)$ configurations of the Iwasaki gauge action at $\beta=2.7$ which has a similar lattice spacing $a=0.0904(74)$.  The configuration numbers  are restricted to 80 in all cases. To reduce lattice artifact monopoles, we introduce an additional gauge fixing called Maximally Abelian gauge with subsequenct $U(1)^2$ Landau gauge (MAU12)\cite{Kronfeld:1987ri,Kronfeld:1987vd}.

 Since the Abelian monopole currents satisfy the conservation condition at each site,  they make a closed loop on the dual lattice. Loop length distributions are measured numerically. The results are summarized in Table \ref{looplength}. 
\begin{enumerate}
  \item There exist one long monopole loop both in most quenched and full QCD cases. For example, see the full QCD case in 
Table \ref{looplength}. The total loop number minus the number of the shortest loops is almost always one. 
  \item The partial gauge fixing introduced here to make the vacuum smooth violates global color invariance. The MA gauge enhances the diagonal color components. But as seen from the diagonal color=3 and the off-diagonal color=1 examples,  large difference is not seen after the subsequent $U(1)^2$ fixing is done. 
  \item The length of the long loop in full QCD is a few times larger than those in quenched case. 
  \item The most interesting fact is that above monopole behaviors are almost unaffected by quark and pion masses as long as the dynamical quark effect is taken into account. This is totally unexpected to the present author. Since monopoles are expected to be important in color confinement, this fact strongly suggests that the mechanisms of color confinement and chiral symmetry breaking in QCD are quite different, since the chiral symmetry depends strongly on the bare quark mass.
  \item Monopole density in full QCD is a few times larger than that in quenched QCD as seen from Table\ref{density}. Also in this case there exist essentially no difference among full QCD configurations with different quark masses. 
  \end{enumerate}
  
\begin{figure}[tbp]
  \begin{minipage}[b]{0.5\linewidth} 
    \centering
 \includegraphics[keepaspectratio, scale=0.38]{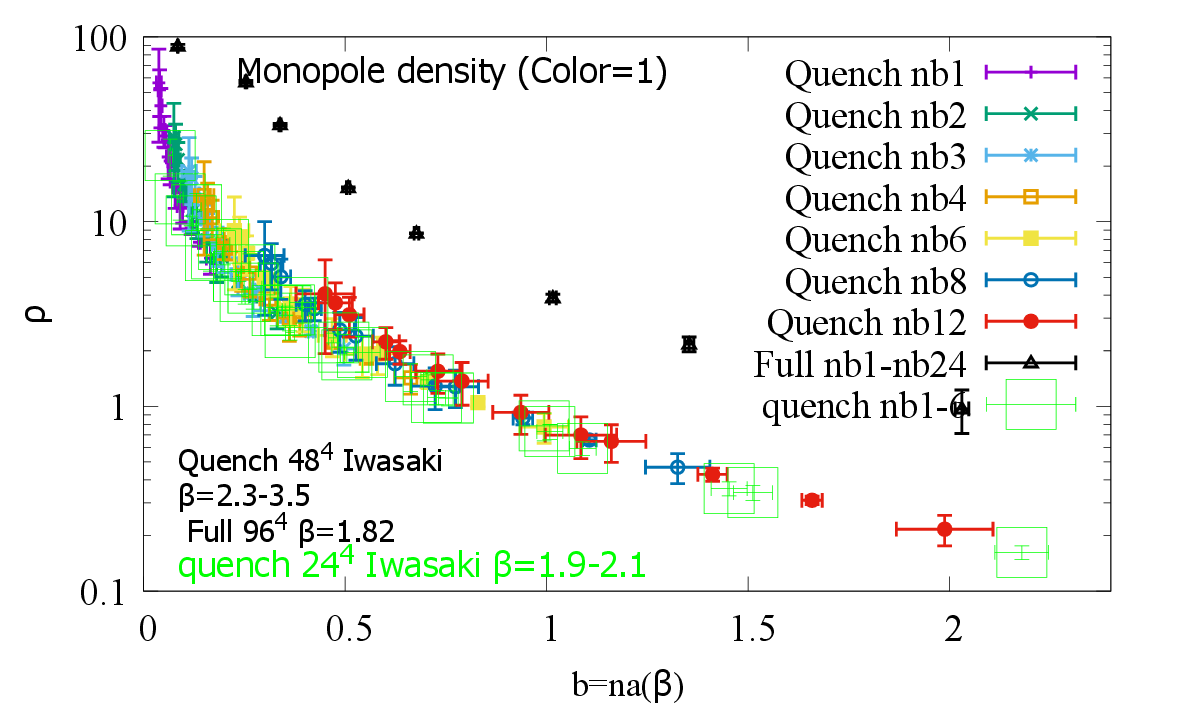}
  \end{minipage}  
  \begin{minipage}[b]{0.5\linewidth}
    \centering
 \includegraphics[keepaspectratio, scale=0.38]{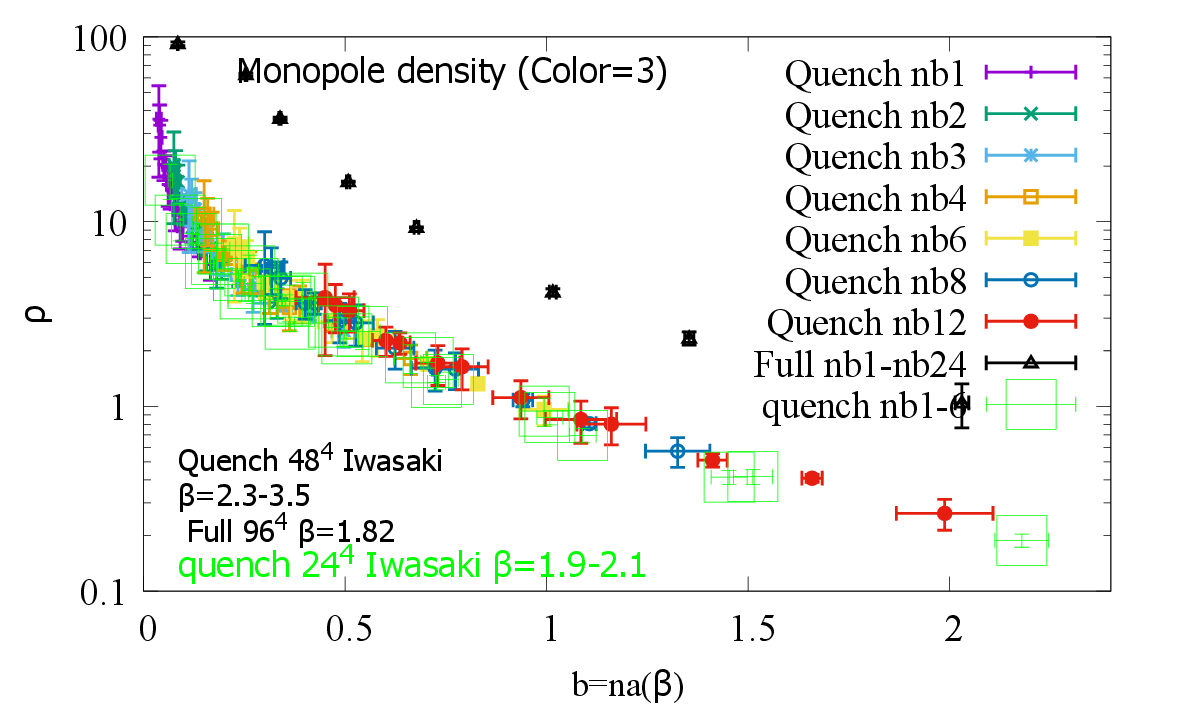}
  \end{minipage} 
\caption{ Density of color=1 (left) and  color=3 (right) monopoles  }
\label{FQ_density}   
\end{figure}
  
 \section{Study of the continuum limit of the Abelian monopoles}  
\label{sec:continuum}
For the purpose of studying the continuum limit of lattice monopoles, the block spin transformation method is very powerful. The idea of the block spin with respect to Abelian monopoles on lattice was first introduced by Ivanenko et al.\cite{Ivanenko:1991wt} and applied to the study obtaining an infrared effective monopole action in Ref.\cite{Shiba:1994db}. The $n$ blocked monopole has a total magnetic charge inside the $n^3$ cube and is defined on
a blocked reduced lattice with the spacing $b=na$. 
The respective magnetic currents for each color are defined as
\begin{eqnarray}
k_{\mu}^{(n)}(s_n)  =  \sum_{i,j,l=0}^{n-1}k_{\mu}(ns_n
     +(n-1)\hat{\mu}+i\hat{\nu}
     +j\hat{\rho}+l\hat{\sigma}), \label{excur}
\end{eqnarray}
where $s_n$ is a site number on the reduced lattice and the color indices are not shown explicitly.
These equations show for example that the relation between $k_{\mu}^{(4)}(s_4)$ and $k_{\mu}^{(2)}(s_2)$ is similar to that between $k_{\mu}^{(2)}(s_2)$ and $k_{\mu}(s)$ and hence the above equation (\ref{excur}) corresponds to the usual block spin transformation.
After the block spin transformation, the number of short lattice artifact  monopole loops decreases while loops having larger magnetic charges appear. For details, see Ref.\cite{Suzuki:2017lco}. 

\subsection{Monopole density}
First we calculate the density of a monopole with color $a$ defined as 
\begin{eqnarray}
\rho^a=\frac{\sum_{\mu,s_n}|k_{\mu}^a(s_n)|}{4\sqrt{8}V_nb^3},\label{eq:Mdensity}
\end{eqnarray}
where $V_n=V/n^4$ is the 4 dimensional volume of the reduced lattice, $b=na(\beta)$ is the spacing of the reduced lattice after $n$ times of the block spin transformation. To reduce lattice artifact monopoles, we have to adopt some partial gauge-fixing smoothing the vacuum. In previous works in $SU(2)$\cite{Suzuki:2017lco} we adopted the maximal center gauge(MCG)\cite{DelDebbio:1996mh,DelDebbio:1998uu} and maximal Abelian gauge(MAG)\cite{Kronfeld:1987ri,Kronfeld:1987vd}. Very clear scaling and gauge independence are seen.  However in $SU(3)$, we adopt only the MAG gauge and the subsequent $U(1)^2$ Landau gauge, where the global $SU(3)$ is broken.  The color-averaged density has been observed in the previous work\cite{Suzuki:2023}, where preliminary analyses of $S(k)$ in $SU(3)$ are made for the first time. To see the color dependence, we here observe both densities of color 1 and color 3 monopoles separately as a typical example of off-diagonal and  diagonal components. They are shown in Figure \ref{FQ_density}.  In the quenched $SU(3)$ case, we adopt $48^4$ Iwasaki gauge action at $\beta=2.3\sim 3.5$, where $a(\beta)=0.04fm\sim 2.0fm$. To study volume dependence, we also study configurations of $24^4$ Iwasaki action at $\beta=1.9\sim 2.1$. For comparison, the full QCD data are also plotted, where the configurations using the 6-APE stout smeared Wilson clover action and Iwasaki gauge action on $96^4$ at $\beta=1.82$ and $m_\pi=146$MeV are cited from the work done by PACS collaboration\cite{Ishikawa:2016}. The densities are shown for various $n=1\sim 24$. Clear scaling and volume independent results are seen in the quenched case. The monopole density in the full QCD is about 3 or 4 times larger than those in the quenched case, although the $b$ dependence looks similar except for very small $b$ region. 
\begin{figure}[htb]
  \begin{minipage}[b]{0.5\linewidth} 
    \centering
 \includegraphics[keepaspectratio, scale=0.38]{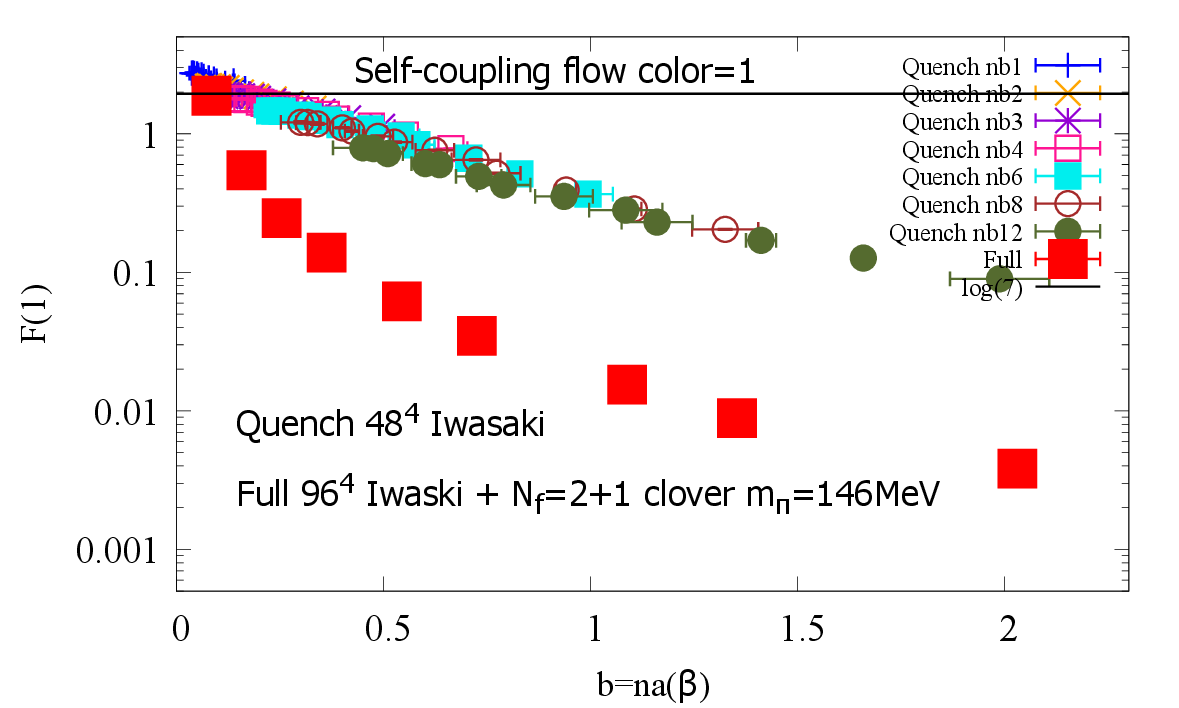}
  \end{minipage}
  \begin{minipage}[b]{0.5\linewidth}
    \centering
 \includegraphics[keepaspectratio, scale=0.38]{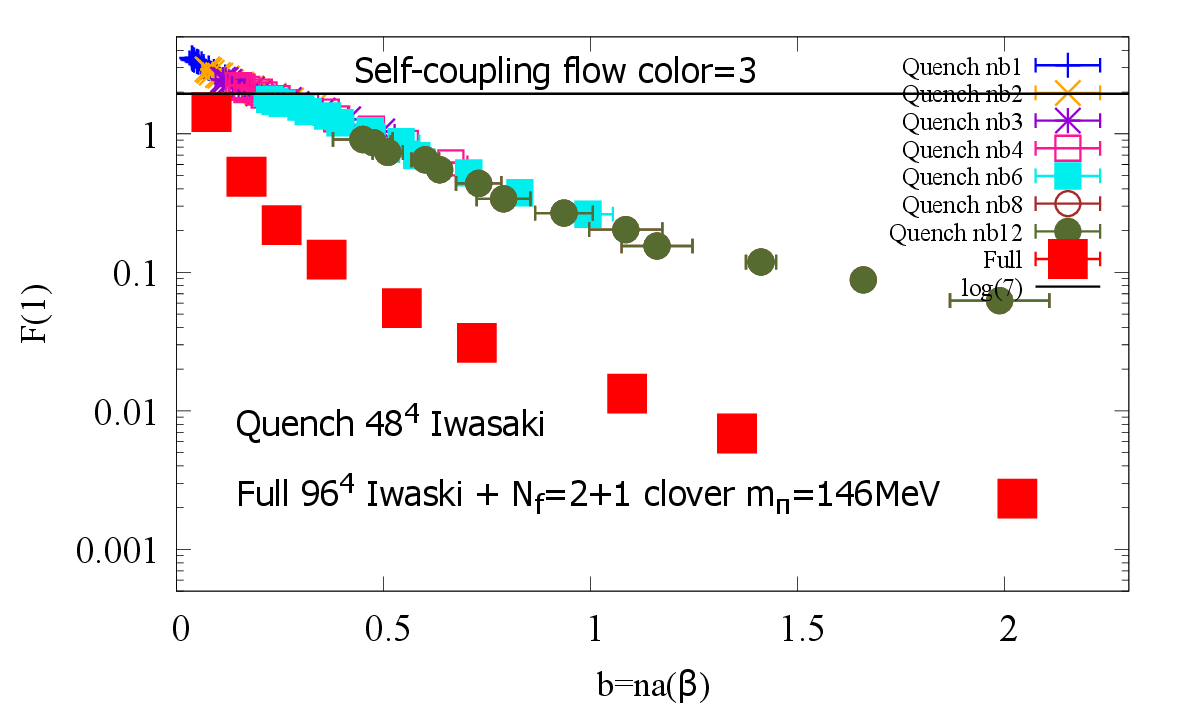}
  \end{minipage}  
\caption{Self-coupling in quenched QCD for the color=1 (left) and for the color=3 (right). }
\label{F1FQ}  
\end{figure}
\begin{figure}[tb]
  \begin{minipage}[b]{0.5\linewidth} 
    \centering
 \includegraphics[keepaspectratio, scale=0.38]{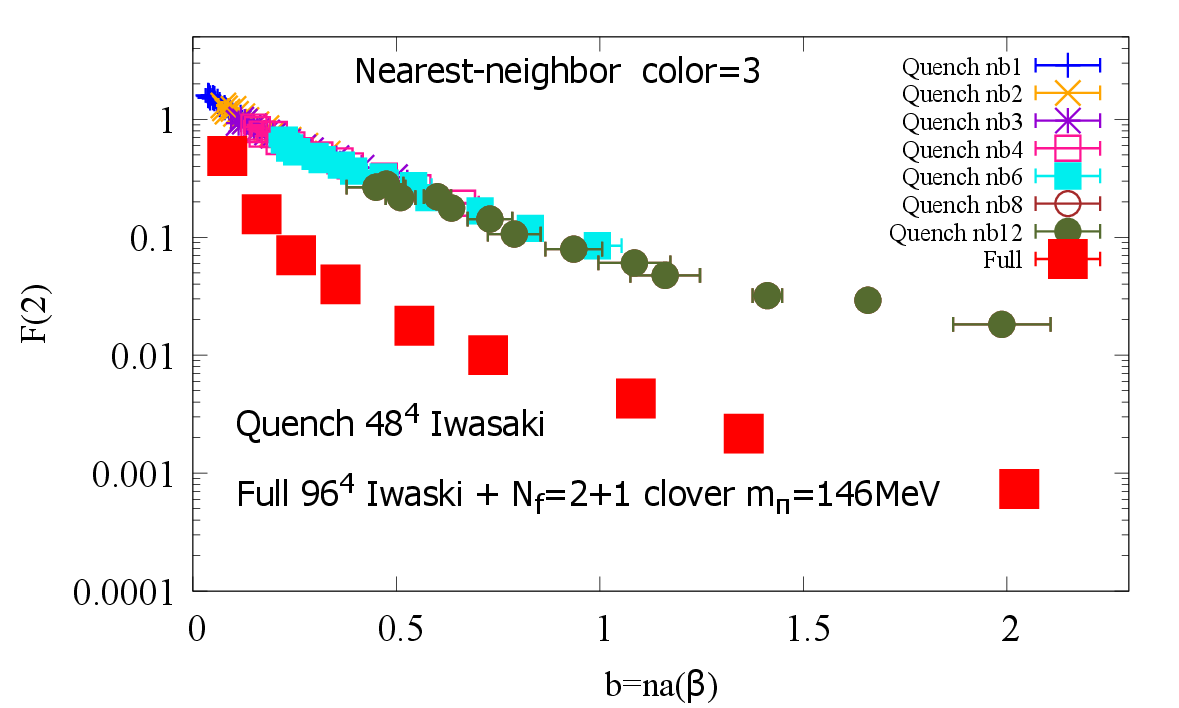}
  \end{minipage}
  \begin{minipage}[b]{0.5\linewidth}
    \centering
 \includegraphics[keepaspectratio, scale=0.38]{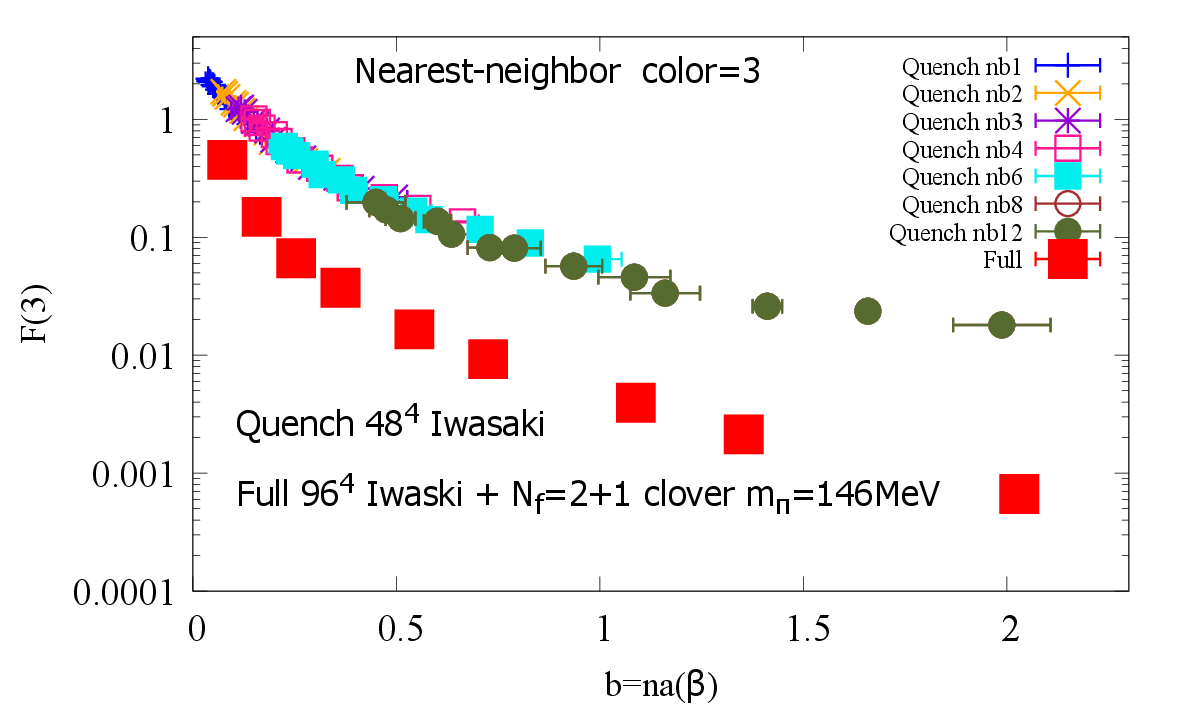}
  \end{minipage} 
\caption{Nearest neighbor-coupling $F(2)$ (left) and $F(3)$ (right) for the color=3. }
\label{F23FQ}  
\end{figure}

\begin{figure}[tb]
  \begin{minipage}[b]{0.5\linewidth} 
    \centering
 \includegraphics[keepaspectratio, scale=0.38]{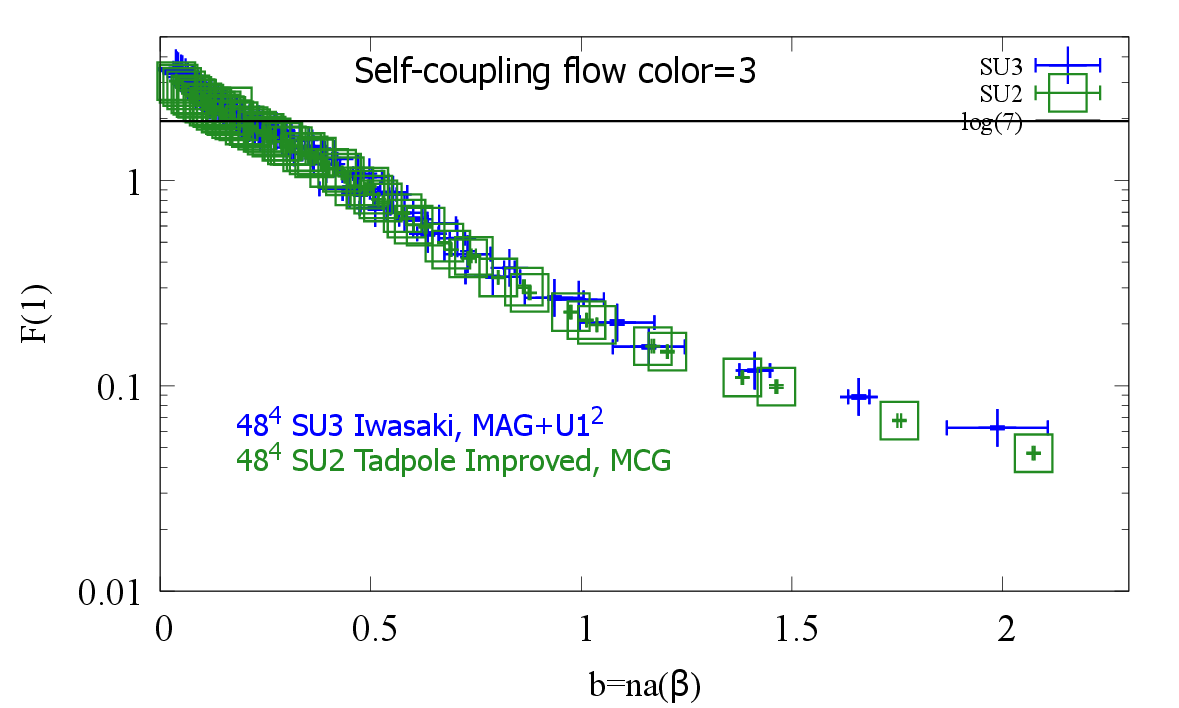}
  \end{minipage}
  \begin{minipage}[b]{0.5\linewidth}
    \centering
 \includegraphics[keepaspectratio, scale=0.38]{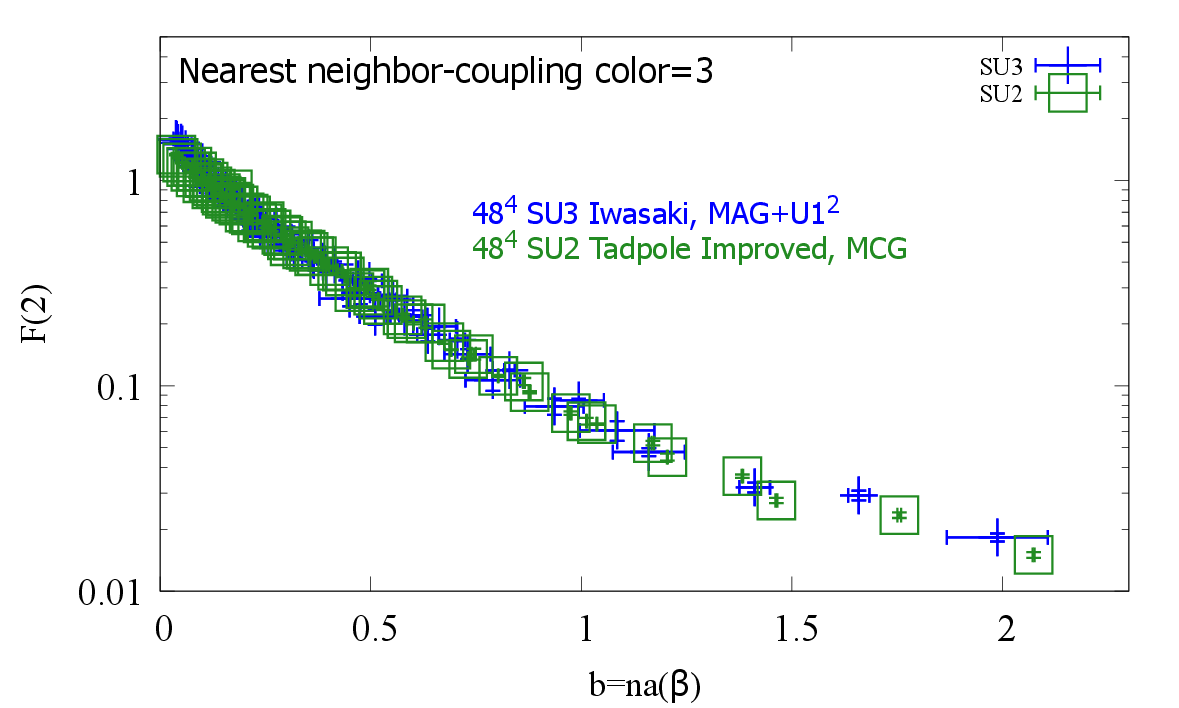}
  \end{minipage}  
\caption{Comparison between $SU(2)$ and $SU(3)$. Self $F(1)$ (left) and neighbor-coupling $F(2)$ (right) for color=3. $SU(2)$ data are from Ref.\cite{Suzuki:2017zdh}.}
\label{SU2SU3}     
\end{figure}

\subsection{The infrared effective monopole action}
It is important to know the dual magnetic theory from QCD. But contrary to  
the compact QED\cite{Villain:1975,Janke:1986,Banks:1977,Peskin:1978,Frolich:1986,Smit:1991,Shiba:1994pu,Jersak:1985}, the exact dual transformation cannot be done in QCD. However,  it is possible  to derive numerically the infrared effective monopole action   
$S(k)$ defined as 
\begin{eqnarray*}
e^{-{\cal S}[k]}=\int DU{\psi}e^{-S(U)}
\prod_{a,\mu,s} \delta(k_{\mu}^a(s)+\frac{1}{4\pi}\epsilon_{\mu\nu\rho\sigma}
\partial_{\nu}\bar{\theta}_{\rho\sigma}^a(s+\hat{\mu})),
\end{eqnarray*}
where $S(U)$ is the QCD action. The effective action for blocked monopoles $S(k^{(n)})$ is defined similarly. Making use of the inverse Monte-Carlo method first proposed by Swendsen\cite{Swendsen:1984} and extended to closed monopole currents by Shiba and Suzuki~\cite{Shiba:1994db}, we get the renormalization flow of infrared effective monopole actions as ${\cal S}[k]\to {\cal S}[k^{(1)}]\to {\cal S}[k^{(2)}] \ldots $. The extensive studies in quenched $SU(2)$ case\cite{Kato:1998ur,Fujimoto:2000,Chernodub:2000,Chernodub:2003,Chernodub:200308} show that interactions between two monopole currents being not far apart are dominant except for small $b<0.5fm$, so that we adopt here first 10 two-point interactions $S[k] = \sum_{i}^{10} F(i) S_i[k]$. For example, $S_1(k)=\sum_{s,\mu} (k_\mu(s))^2$,  $S_2(k)=\sum_{s,\mu} k_\mu(s)k_\mu(s+\mu)$ and $S_3(k)=\sum_{s,\mu\neq\nu} k_\mu(s)k_\mu(s+\nu)$. Others are shown in Table VI in the previous work\cite{Suzuki:2023}. Since we deal with only two-point interactions between monopoles of the same color, we omit the color 
index $a$.  For details of the inverse Monte-Carlo method, see Ref.\cite{Suzuki:2017zdh}.

\begin{figure}[tb]
  \begin{minipage}[b]{0.5\linewidth} 
    \centering
 \includegraphics[keepaspectratio, scale=0.38]{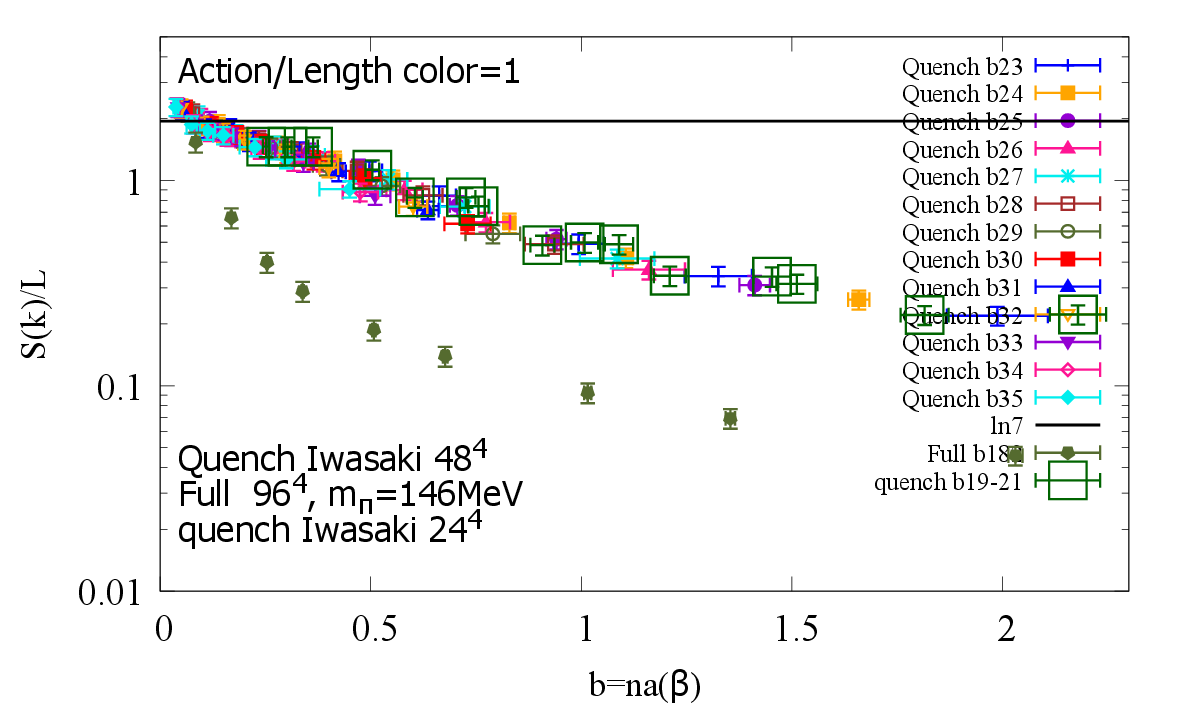}
  \end{minipage}
  \begin{minipage}[b]{0.5\linewidth}
    \centering
 \includegraphics[keepaspectratio, scale=0.38]{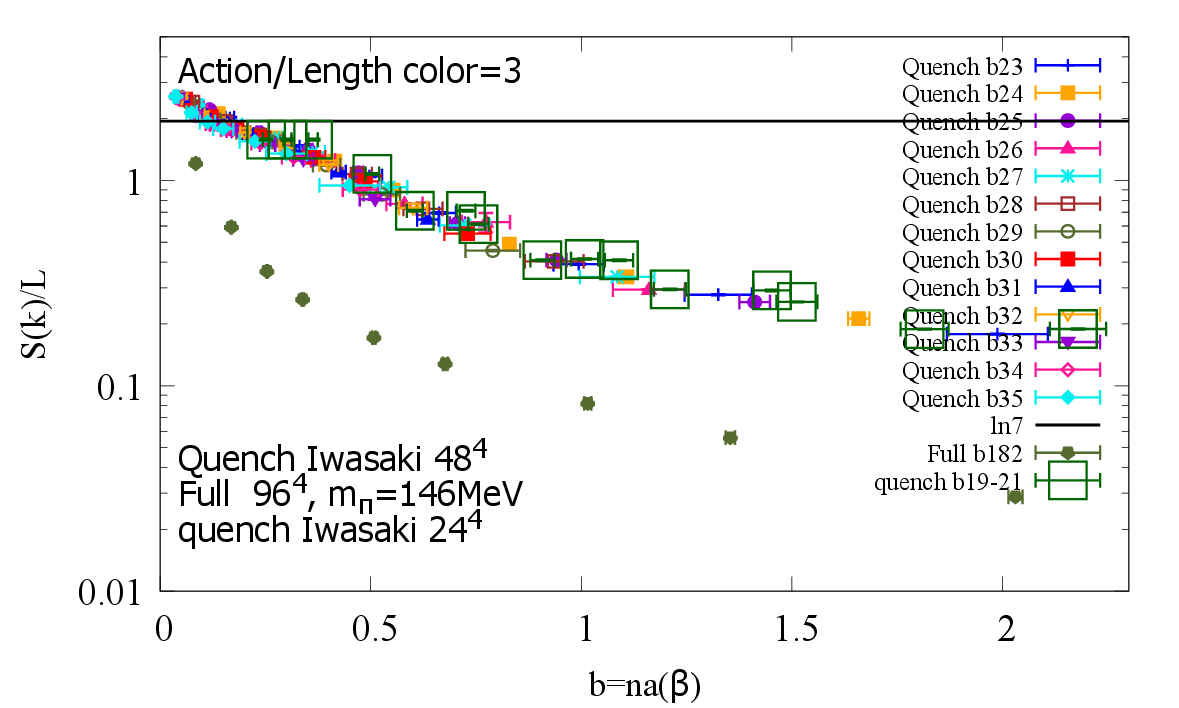}
  \end{minipage}  
\caption{Action/Length ($S(k)/L$)  of the color=1 (left) and  the color=3 (right) monopoles}
\label{AL}  
\end{figure}

The results are as follows:
\begin{enumerate}
  \item Self ($F(1)$) and two nearest neighbor ($F(2)$, $F(3)$) coupling constants both in quenched and full cases are shown in Figure \ref{F1FQ} and in Figure \ref{F23FQ}.     
These coupling constants $F(i)$ are originally a two-point function of $a(\beta)$ and the number of blocking times $n$, but actually they are found to be a function of $b=na(\beta)$ alone. The beautiful scaling behaviors in the quenched case suggest that we are seeing the continuum limit, since $n\to\infty$ corresponds to $a(\beta)\to 0$ for fixed $b$. The coupling constants in full QCD are always lower than those in the quenched data. Moreover the full QCD data show stronger decreasing behaviors as $b$ becomes larger, although the scaling is not yet checked in the full QCD case  taken at $\beta=1.82$ alone. 
  \item Both quenched $SU(2)$\cite{Suzuki:2017zdh} and $SU(3)$ results are found to be very similar as shown in Figure \ref{SU2SU3}, although different gauge-fixings are adopted. But note that both deal with only a $U(1)$ effective monopole action. 
  \item The results on $24^4$ and $48^4$ lattices are found to be almost the same. Volume dependence is not seen as seen from Figure \ref{AL} shown later.
\end{enumerate}

\begin{figure}[tb]
  \begin{minipage}[b]{0.5\linewidth} 
    \centering
 \includegraphics[keepaspectratio, scale=0.4]{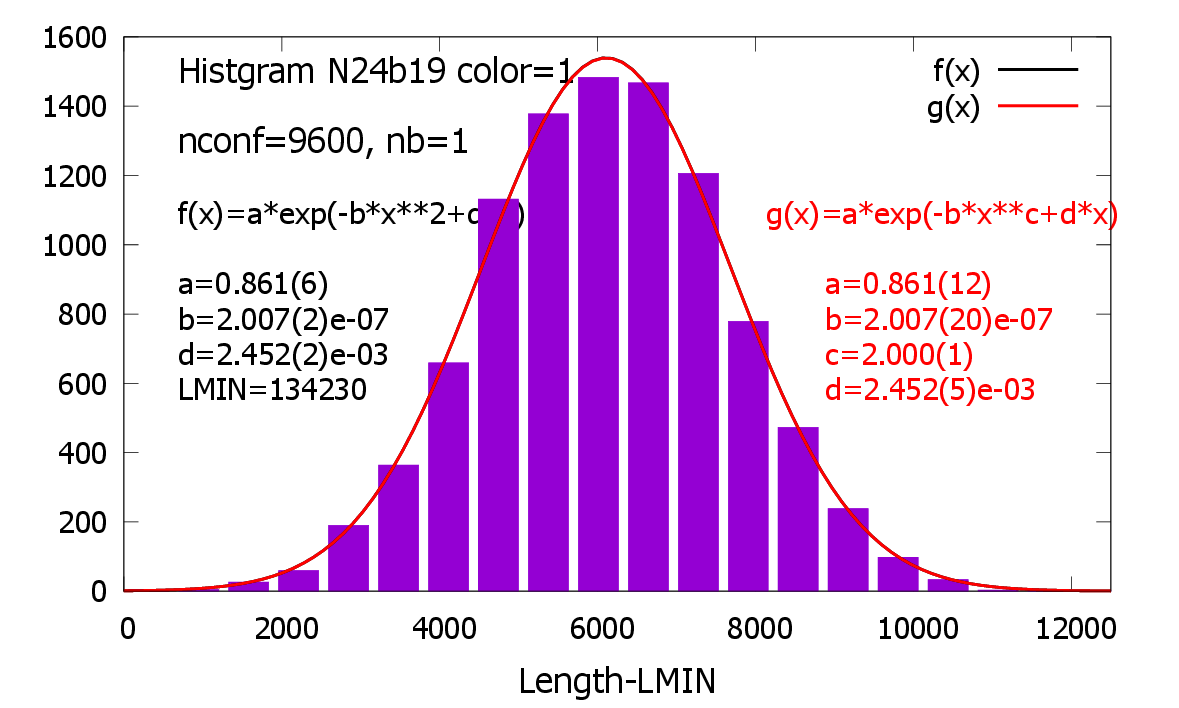}
  \end{minipage}
  \begin{minipage}[b]{0.5\linewidth}
    \centering
 \includegraphics[keepaspectratio, scale=0.4]{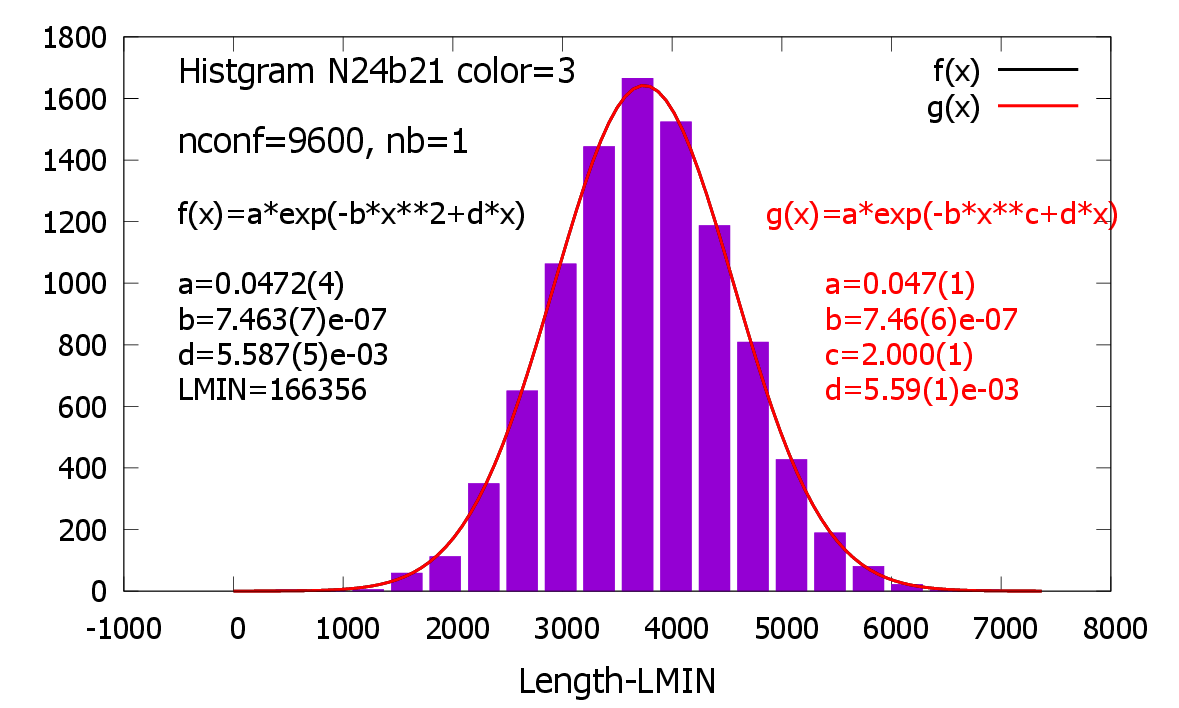}
  \end{minipage}  
\caption{Loop length histogram of percolating monopoles on  $24^4$ of color 1  at $\beta=1.9$ (left) and of color 3 at $\beta=2.1$}
\label{MD}  
\end{figure}

\begin{figure}[tb]
  \begin{minipage}[b]{0.5\linewidth} 
    \centering
 \includegraphics[keepaspectratio, scale=0.38]{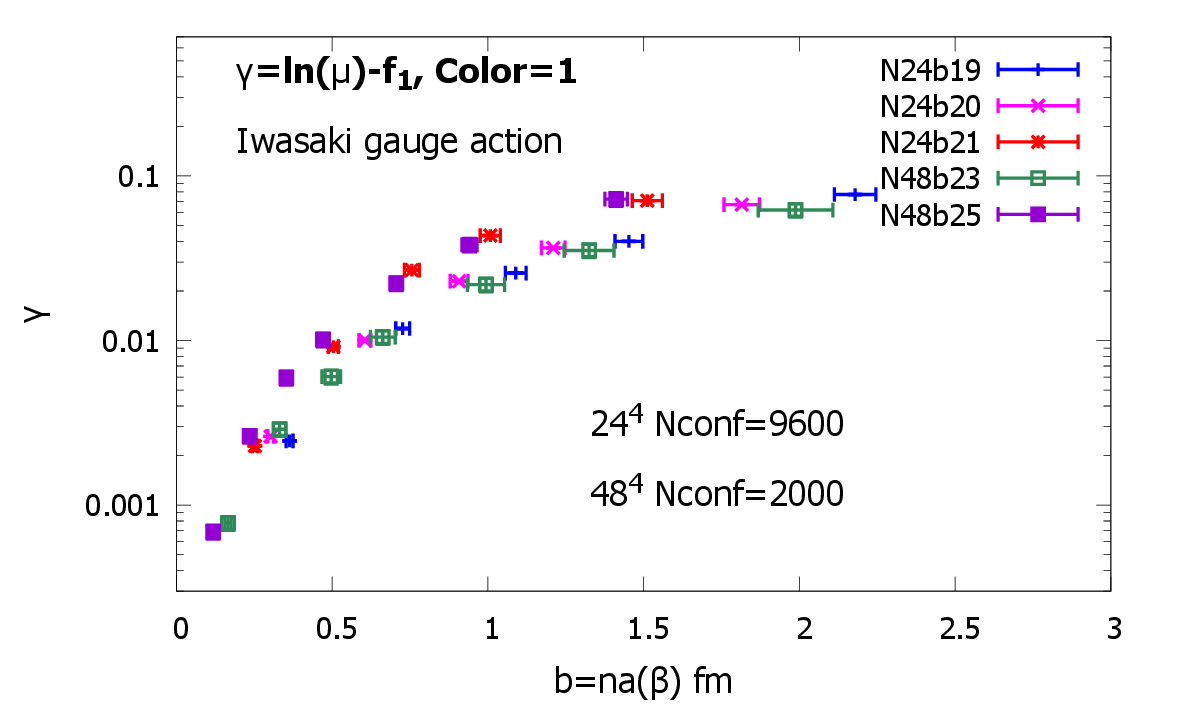}
  \end{minipage}
  \begin{minipage}[b]{0.5\linewidth}
    \centering
 \includegraphics[keepaspectratio, scale=0.38]{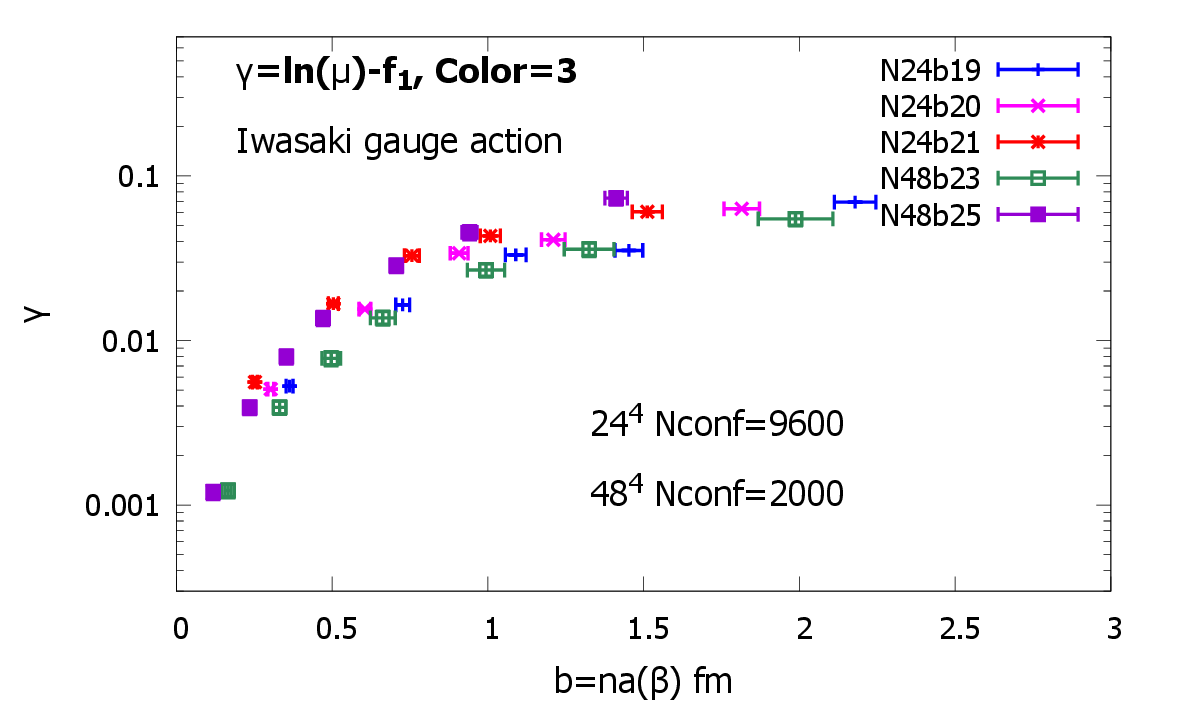}
  \end{minipage} 
\caption{Entropy-Energy $\gamma=\textrm{ln}\mu-f_1$  of the color=1 (left) and the color 3 (right) components} 
\label{gamma} 
  
\end{figure}

\subsection{Energy entropy balance and monopole condensation}
Now that infrared effective monopole actions are derived numerically as above in quenched and full QCD, let us go to the discussion about the monopole condensation on the basis of the energy-entropy balance  as a clue  to color confinement. 

    Since the monopole currents are conserved, they form closed loops on the dual links. The distribution of the monopole loops are known to be composed of a very long percolating loop running over the whole volume and some short ultraviolet loops. See \cite{Chernodub:200308} in $SU(2)$ and  in $SU(3)$ see Table\ref{looplength}.  Figure \ref{AL} shows that the total monopole action is proportinal to the percolating monopole length $L$. The entropy of monopole currents of length $L$ can be expressed essentially as $\textrm{ln}(\mu^L)$, where $\mu$ is some positive number. When the monopoles obey the random walk, $\mu=7$ in four dimensions as seen in compact QED\cite{Smit:1991}. Here since we are considering also the blocked monopoles, $\mu$ is not fixed theoretically. The monopole partition function is hence well approximated by the sum of the monopole loop distribution function $D(L)$ over the loop lengths of the monopoles:
\begin{eqnarray}
D(L)\propto \mu^L e^{-S(L)}=e^{\gamma(b) L},
\end{eqnarray}
where $\gamma(b)=\textrm{ln}\mu-f_1(b)$ and $S(L)=f_1(b)L$. When the lattice volume is finite, the monopole length $L$ cannot become infinite and so there must exist a certain cut depending on the lattice volume. We choose it following Ref.\cite{Chernodub:200308}  as follows:
\begin{eqnarray}
D(L)=e^{-\alpha(b,V) L^{\eta}+\gamma(b) L}, \label{DL}
\end{eqnarray}
where $\alpha, \eta, \gamma$ are some parameters. In (\ref{DL}), there can exist a power law term coming from the ultraviolet short loops, but it is negligible as checked in Ref.\cite{Chernodub:200308}. The function (\ref{DL}) can reproduce well the actual distribution of percolating loops as shown in Figure \ref{MD} where 9600 monopole configurations on $24^4$ of color 1 at $\beta=1.9$ and of color 3 at $\beta=2.1$ are used as typical examples. The parameter $\eta$ is found  to be almost 2.0 showing the Gaussian distribution. The same situations are seen in all other cases irrespectively of color, $\beta$ and number of blocking times $n$ on $24^4$.  Hence we fix $\eta=2.0$ hereafter. The peak of the distribution is then given by 
\begin{eqnarray} 
L_{max}=\gamma(b)/(2\alpha(b,V)).
\end{eqnarray}
The monopole density $\rho$ showing the scaling as seen from Figure \ref{FQ_density} and the definition of the monopole density (\ref{eq:Mdensity}) give  
\begin{eqnarray}
\rho \propto L_{max}/V,
\end{eqnarray}
so that we see $\alpha(b,V)=A(b)/V$. These parameters are fixed from the analyses of loop distributions as seen from an example of Figure \ref{MD}. However, on larger lattices,
it is difficult to adopt so many configurations. Hence on $48^4$, we first generate 2000 configurations as usual. Assuming the Gaussian distribution, we can fix the parameters as 
\begin{eqnarray}
\alpha=\frac{1}{2}\frac{1}{<L^2>-<L>^2},\ \ \ \ \gamma=\frac{<L>}{<L^2>-<L>^2},\label{AG}
\end{eqnarray}
where $<L^2>$ and $<L>$ are determined from the histograms. To evaluate the errors of the parameters, we adopt the bootstrap method\cite{Chernodub:200308} by generating randomly  2000 sets of 2000 configurations. In Figure \ref{gamma}, we plot $\gamma(b)=\textrm{ln}\mu-f_1(b)$.  $\gamma(b)$ looks to be a function of $b=na(\beta)$ alone for large $b$ and  almost volume independent. Note that these data of $\gamma(b)$ are all positive indicating that the entropy of the monopole loops dominates over the energy. Hence monopole condensation is realized for all $b=na(\beta)$ regions in quenched QCD. In the case of full QCD, the present number of configurations are too few to analyze the monopole loop distribution clearly. However the distribution of long loops is consistent with the above Gaussian form having a positive $\gamma$.

\section{Concluding remarks}
If VNABI exists in QCD, there appear magnetic $U(1)_m^8$ kinematical symmetries corresponding to 8 Abelian conserved monopole currents. It is explicitly shown that if the  magnetic $U(1)_m^8$ are broken spontaneously,  non-Abelian color confinement is explained by the Abelian dual Meissner effect. Next it is shown in the framework of lattice QCD  that clear scaling behaviors especially in quenched QCD with respect to the monopole density and the infrared effective action are observed.  Finally analyzing the monopole loop distribution, we prove that the monopole condensation occurs for all $b$ regions discussed here due to the energy-entropy balance. These results show that the new Abelian monopoles corresponding to the violation of non-Abelian Bianchi identity are very hopeful candidates for color confinement.  It is also interesting to check if the energy-entropy arguments work also in explaining the confinement-deconfinement transition in finite-temperature QCD.
Analyses of the effective action of the spacelike monopoles projected onto the three dimensional time-slice show strongly that the entropy-energy balance is the mechanism of the confinement-deconfinement transition in $SU(2)$\cite{CIS:2005}. Similar studies are expected in quenched and dynamical $SU(3)$.  

Although we have discussed only infrared effective monopole actions, various infrared effective theories of QCD can be derived from the above Abelian monopole actions\cite{Chernodub:1999}.   
  The monopole action composed of quadratic interactions alone corresponds to the dual Abelian Higgs (DAH) model in the London limit, where the DAH model\cite{Suzuki:1988,Maedan:1989} is given by
\begin{eqnarray}
L_{DAH}=\frac{1}{4g_m^2}(\partial_\mu B_\nu-\partial_\nu B_\mu)^2 
 +\frac{1}{2}|(\partial_\mu+iB_\mu)\Phi|^2+\lambda(|\Phi|^2
-\eta^2)^2.
\end{eqnarray}
Here $B_\mu$ is a dual Abelian gauge field and $\Phi$ is a complex scalar monopole field. In the London limit $\lambda\to\infty$, the radial part of the monopole field is frozen, that is, $|\Phi|=\eta$. In this limit, 
the mass of the scalar field becomes infinite. But from the analyses of the penetration and the coherence lengths, we expect that the masses of the dual gauge and the scalar fields are roughly equal both in $SU(2)$\cite{Suzuki:2009} and $SU(3)$\cite{IHS:2022} cases. Hence higher-order interactions such as
four- and six-point monopole interactions are necessary. Actually from the extensive studies in quenched $SU(2)$ QCD, we know such higher-order interactions become relevant for $b<0.5fm$ regions\cite{Chernodub:2000}.
Previous results on $24^3\times 4$\cite{IHS:2022} in quenched QCD suggest that the vacuum type in $SU(3)$ is the weak type I (dual) superconductor near the border. Also the mass of the dual gauge boson determined from the penetration length is about 1GeV.  If the quenched results are not changed much in full QCD at the physical point, both axial and scalar bosons are predicted to have a mass roughly around 1 GeV.  To find such new scalar and axial-vector bosons experimentally is crucial to the confinement picture proposed here. Note that both the scalar and the axial vector bosons are color-singlet and eightfold degenerate exactly.  
In addition to the dual Abelian Higgs model, the hadronic string model can be derived exactly from the Abelian monopole action\cite{Chernodub:1999} as naturally expected.  

These results obtained in this work are consistent with the perfect Abelian and monopole dominances with respect to the string tension showing the area-law behaviors of non-Abelian Wilson loops\cite{Suzuki:2009,IHS:2022,Suzuki:2023}.

\acknowledgments
The numerical simulations of this work were done  using High Performance Computing resources at Research Center for Nuclear Physics  (RCNP) of Osaka University.  The author would like to thank  RCNP for their support of computer facilities. This work is finacially supported by JSPS KAKENHI Grant Number JP19K03848. This work is supported by the JLDG constructed over the SINET5 of NII. The author thanks the PACS-CS and the PACS Collaborations
 for making their 2 +1 flavour configurations available via the International Lattice Data Grid (ILDG).

\end{document}